\documentclass[aps,twocolumn,prd,superscriptaddress,showpacs,floatfix,preprintnumbers,nofootinbib]{revtex4}   

\usepackage{graphicx}\usepackage{natbib}\usepackage{amsmath}
\usepackage{epsfig}   
\usepackage[dvips]{color}
\definecolor{Red}{named}{Red}
\definecolor{Blue}{named}{Blue}
\definecolor{Black}{named}{Black}

\begin{document}   
\preprint{MPP-2009-203}   

\title{Detecting the QCD phase transition in the next Galactic supernova neutrino burst}

\author{Basudeb Dasgupta}   
\affiliation{Max-Planck-Institut f\"ur Physik (Werner-Heisenberg-Institut), F\"ohringer Ring 6, 80805 M\"unchen, Germany}  

\author{Tobias Fischer}   
\affiliation{Department of Physics, University of Basel, Klingelbergstr. 82, CH-4056 Basel, Switzerland} 

\author{Shunsaku Horiuchi} 
\affiliation{Institute for the Physics and Mathematics of the Universe, University of Tokyo, 5-1-5, Kashiwanoha, Kashiwa, Chiba 277-8582, Japan}
\affiliation{Center for Cosmology and Astro-Particle Physics, Ohio State University, Columbus, Ohio 43210}

\author{Matthias~Liebend{\"o}rfer}   
\affiliation{Department of Physics, University of Basel, Klingelbergstr. 82, CH-4056 Basel, Switzerland}

\author{Alessandro Mirizzi}   
\affiliation{II Institut f\"ur Theoretische Physik, Universit\"at Hamburg, Luruper Chaussee 149, 22761 Hamburg, Germany}

\author{Irina Sagert}   
\affiliation{Institut f\"ur Theoretische Physik, Goethe Universit\"at, Max-von-Laue-Str. 1, 60438 Frankfurt am Main, Germany}
\affiliation{Institut f\"ur Theoretische Physik, Ruprecht-Karls-Universit\"at, Philosophenweg 16, 69120 Heidelberg, Germany}

\author{J\"urgen Schaffner-Bielich}   
\affiliation{Institut f\"ur Theoretische Physik, Ruprecht-Karls-Universit\"at, Philosophenweg 16, 69120 Heidelberg, Germany}

\date{\today}   


\begin{abstract}   
Predictions of the thermodynamic conditions for phase transitions at high baryon densities and large chemical potentials are currently uncertain and largely phenomenological. Neutrino observations of core-collapse supernovae can be used to constrain the situation. Recent simulations of stellar core-collapse that include a description of quark matter, predict a sharp burst of $\bar{\nu}_e$ several hundred milliseconds after the prompt $\nu_e$ neutronization burst. We study the observational signatures of that $\bar{\nu}_e$ burst at current neutrino detectors -- IceCube and Super-Kamiokande. For a Galactic core-collapse supernova, we find that signatures of the QCD phase transition can be detected, regardless of the neutrino oscillation scenario. The detection would constitute strong evidence of a phase transition in the stellar core, with implications for the equation of state at high matter density and the supernova explosion mechanism.
\end{abstract}   


\maketitle       

\section{Introduction} 

The majority of stars more massive than \(8\) M\(_\odot\) end their lives as core-collapse supernovae (SNe), with explosive kinetic energies \(\sim10^{51}\) erg. The explosion mechanism is most likely related to the revival of the stalled bounce shock -- the shock that forms when the collapsing iron core reaches nuclear densities, and stalls on its way out due to continuous energy loses via neutrino emission and dissociation of heavy nuclei. The detailed explosion mechanism has not been unambiguously identified yet. Given the surrounding matter envelope is opaque for photons, neutrinos have been highly sought after for studying the physical conditions, dynamics of the collapse, and the SN mechanism. Indeed, current neutrino detectors are expecting high statistics from the next Galactic SN, and are on the verge of detecting the diffuse neutrino background from all past SNe~\cite{Malek:2002ns,Ando:2004hc,Horiuchi:2008jz}. These would enable the explosion mechanism to be tested~\cite{Woosley:2002zz,Totani:1997vj,Dighe:2003be,Halzen:2009sm}, as well as neutrino mixing parameters to be constrained (see, e.g., Ref.~\cite{Dighe:2008dq}). Moreover, neutrinos could potentially reveal new physics operating deep in the stellar core.

Phase transitions have long been investigated in the context of SNe~\cite{Takahara:1988yd} and neutron stars~\cite{Migdal:1974aa}. By including prescriptions of additional matter in the equation of state (EoS), many simulations predict the appearance of strange matter in the form of hyperons, a kaon condensate, or quark matter at supernuclear densities~\cite{Migdal:1974aa,Pons:2001,Drago:2008tb}. Bose condensates of pions have also been studied. In this paper we concentrate on recent developments in quark-hadron phase transitions. Initial attempts were based on general relativistic hydrodynamics using a parametrized equation of state~\cite{Takahara:1988yd}, discussing possible relations to the neutrino spectrum of SN1987A~\cite{Hirata:1988}. Applying a more sophisticated equation of state, the formation of a strong second shock wave was found as a direct consequence of the phase transition~\cite{Gentile:1993ma}. However, these models could not predict the post-bounce neutrino signal due to the lack of neutrino transport. Phase transitions were also investigated in the context of very massive ($\sim100 M_\odot$) progenitor stars, where the time until black hole formation is shortened due to the softening of the equation of state during the quark-hadron phase transition~\cite{Thorsson:1993bu,Pons:2001,Nakazato:2008su}. Recently, applying general relativistic radiation hydrodynamics based on three-flavor Boltzmann neutrino transport in spherically-symmetric simulations of low- and intermediate-mass progenitor stars, combined with the MIT bag model for the description of strange quark matter, the formation of a strong second shock wave was confirmed~\cite{Sagert:2008ka}. The second shock accelerates at the surface of the protoneutron star and merges with the bounce shock, triggering an explosion, where otherwise no explosion could be obtained. As the second shock crosses the neutrinospheres, a second burst of neutrinos is released. Since the degeneracy is lifted in the newly shocked material and the electron fraction must be increased, the second burst is dominated by $\bar{\nu}_e$.

\begin{figure}[!ht]
\centering
\epsfig{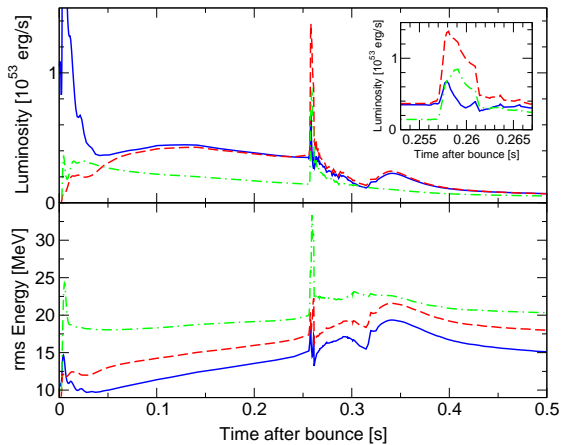}
\caption{Neutrino luminosities and rms neutrino energies as functions of time after bounce, sampled at 500 km radius in the comoving frame, for a \(10\) M\(_\odot\) progenitor star as modeled in \cite{Sagert:2008ka}: \(\nu_e\) in solid (blue), \(\bar\nu_e\) in dashed (red), and \(\nu_{\mu/\tau}\) in dot-dashed (green). In contrast to the deleptonization burst just after bounce (\(t \sim 5\) ms) the second burst at \(t \sim 257-261\) ms is associated with the QCD phase transition. The inset shows the second burst blown up. 
\label{simulation}} 
\end{figure}


Current neutrino detectors like IceCube and Super-Kamiokande are mostly sensitive to SN ${\bar\nu}_e$'s through inverse beta decay reactions (\({\bar\nu}_e p \to n e^{+}\)) in the detector medium, while $\nu_e$'s are detected by the subleading elastic scattering channel. Encouragingly, the second burst shows up most prominently in ${\bar\nu}_e$'s, and its detection appears feasible already. This contrasts with the deleptonization burst which has been widely explored and proposed as a crucial tool to determine the flavor oscillation effects on the SN neutrino signal~\cite{Kachelriess:2004ds,Duan:2007sh}; since the deleptonization burst consists of $\nu_e$'s, its detection requires larger neutrino experiments. If the \({\bar\nu}_e\) burst associated with a phase transition is observed, it would constitute a strong signature for the presence of a phase transition in the SN core, with important implications for new physics and the SNe explosion mechanism. We consider this intriguing possibility and investigate the capabilities of the largest existing neutrino detectors to unambiguously measure the ${\bar\nu}_e$ burst from a Galactic SN (assumed at a distance of 10~kpc unless stated otherwise).

\section{Numerical model}

We employ the neutrino emissions of Ref.~\cite{Sagert:2008ka}, which simulated the core-collapse of massive stars using a quark matter equation of state based on the widely used MIT bag model. The main physical uncertainty in modeling the QCD phase transition is the critical density at which the transition sets in. In the simulations of Ref.~\cite{Sagert:2008ka}, this is determined by the bag constant and the strange quark mass, chosen such that a quark-hadron transition sets in during the early post-bounce phase, close to nuclear saturation density.

The neutrino emission of the reference simulation of Ref.~\cite{Sagert:2008ka} is shown in Fig.~\ref{simulation}. The ordinary deleptonization burst is seen at \(t\simeq5\)~ms after bounce. At about the time when the quark matter core is created (\(t\simeq 260\)~ms for the simulation shown) a strong second shock wave forms, which crosses the neutrinospheres in a few milliseconds and releases a second burst comprising of ${\bar\nu}_e$'s. The rise-time is a few ms, being related to the shock crossing the neutrinospheres. The duration is \(<4\) ms and the time-integrated energetics is approximately \(5\times10^{50}\) erg in \(\bar{\nu}_e\)'s, $\sim 1$\% of the total energetics. Additionally, the burst is accompanied by a sharp rise in the neutrino average energies. The fluctuations in the neutrino luminosity shortly after the second burst are due to the appearance of an accretion shock when cooling at the neutrinospheres leads to the fallback of the innermost ejecta. The accretion rate and position of the accretion shock varies with time and modulates the luminosity until it settles to a quasi-stationary state on a time scale of about 100 ms.

A second set of simulations was performed using a larger bag constant which results in a higher critical density. The post-bounce time for the second burst is 448 ms (see Table I of Ref.~\cite{Sagert:2008ka}), but otherwise, the burst duration ($\sim 4$ ms) and energetics ($\sim 10^{50}$ erg in $\bar{\nu}_e$'s) are similar to those shown in Fig.~\ref{simulation}. The timing of the second burst after the deleptonization burst contains correlated information about the quark-hadron equation of state, the critical conditions for the quark-hadron phase transition, and the progenitor model.

\section{Neutrino flavor conversions}

In order to determine the SN $\bar{\nu}_e$ signal observed at Earth, one must take into account flavor conversions occurring during propagation. In general, neutrino oscillation effects vary during the post-bounce evolution. In particular, collective flavor conversions in the SN~\cite{Sigl:2009cw} are forbidden during the second burst. From Fig.~\ref{simulation}, one realizes that during this phase the $\bar{\nu}_e$ flux is strongly enhanced with respect to the non-electronic species $\nu_x$ and $\bar{\nu}_x$ (where $x=\mu,\tau$) while the $\nu_e$ flux is strongly suppressed. In this condition, the conservation of the lepton number prevents the collective $\nu_e{\bar\nu}_e \to \nu_x{\bar\nu}_x$ pair conversions~\cite{Hannestad:2006nj}. Therefore, only Mikheyev-Smirnov-Wolfenstein (MSW) flavor conversions occur while the neutrinos propagate through the stellar envelope~\cite{Dighe:1999bi}.

In inverted mass hierarchy (IH: $m_3 < m_1 < m_2$, where $m_i$ is the neutrino mass), MSW matter effects in the SN envelope are characterized in terms of the level-crossing probability $P_H$ of antineutrinos, which is in general a function of the neutrino energy and of the  1--3 leptonic mixing angle $\theta_{13}$~\cite{Dighe:1999bi}. In the following, we consider two limits, namely $P_{H} \simeq 0$ when $\sin^2 \theta_{13} \gtrsim 10^{-3}$ (large) and $P_H \simeq 1$ when $\sin^2 \theta_{13} \lesssim 10^{-5}$ (small). Neglecting for simplicity Earth matter crossing effects, the electron antineutrino flux $F_{\bar{\nu}_e}$ at the Earth surface for IH with small $\theta_{13}$, is given in terms of the primary fluxes $F^0_\nu$ by~\cite{Dighe:1999bi}


\begin{equation}
F_{{\bar\nu}_e} \simeq \cos^2\theta_{12} F^0_{{\bar\nu}_e} +
\sin^2\theta_{12} F^0_{{\bar\nu}_x} \,\ ,
\label{eq:normal}
\end{equation}


where  \(\cos^2\theta_{12} \simeq 2/3\),  with \(\theta_{12}\) being the ``solar'' mixing angle~\cite{Fogli:2008ig}. For IH with large \(\theta_{13}\) one has


\begin{equation}
F_{{\bar\nu}_e} \simeq F^0_{{\bar\nu}_x} \,\ .
\label{eq:invert}
\end{equation}


For normal mass hierarchy (NH: \(m_1 < m_2 < m_3\)) the expression in Eq.~(\ref{eq:normal}) is applicable. The signal is expected to be larger for NH or IH with small \(\theta_{13}\) than that for IH with large \(\theta_{13}\), because \(F^{0}_{{\bar\nu}_e}> F^{0}_{{\bar\nu}_x}\).

Outside the ${\bar\nu}_e$ burst, one finds \(F^{0}_{\nu_e} >F^{0}_{{\bar\nu}_e}>F^0_{\nu_x}\). This flux ordering allows  for both collective and matter flavor transitions in IH. $F_{\bar{\nu}_e}$ is given by Eq.~(\ref{eq:normal}) for large $\theta_{13}$, and Eq.~(\ref{eq:invert}) for small $\theta_{13}$, i.e., effectively the case of large and small $\theta_{13}$ are exchanged relative to the case with no collective effects. For NH collective oscillations are usually unimportant, and we find the same result as in Eq.~(\ref{eq:normal}) for the final flux.

\section{Detection of the QCD neutrino burst}

We analyze the detection of the QCD phase transition induced ${\bar\nu}_e$ burst in two neutrino detectors,  the {km}$^{3}$ ice \u{C}erenkov detector IceCube and the 32 kton water \u{C}erenkov detector Super-Kamiokande. The detectors have distinct advantages -- IceCube is greater in volume, while Super-Kamiokande allows spectral energy reconstruction.


\begin{figure}[t]

\centering

\epsfig{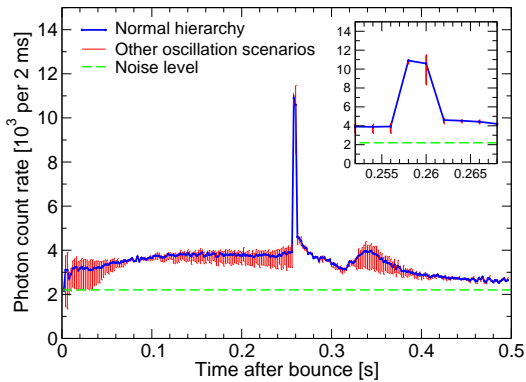}

\caption{Photon count rates at IceCube for the neutrino emission shown in Fig.~\ref{simulation}. The inset shows the second burst blown up, in the same axis units.
\label{IceCube}} 
\end{figure}


\subsection{Detection at IceCube}

To estimate the event rate produced by inverse-$\beta$ decay in the ice, we follow the recent calculation of Halzen and Raffelt~\cite{Halzen:2009sm}. The complete detector will have 4800 optical modules and the data are read out in 1.6384~ms bins, implying a total event rate of 
\begin{equation}
R_{{\bar\nu}_e} = 1860 \,\ \textrm{bin}^{-1} L_{53} \,\ d^{-2}_{10} \,\  \frac{\langle E_{15}^3\rangle }{ \langle E_{15} \rangle^3} ,
\end{equation}
where $L_{53}$ is the $\bar{\nu}_e$ luminosity in units of $10^{53}$ erg s$^{-1}$, $d$ is the distance to the SN in units of 10 kpc, and $E_{15}$ is the $\bar{\nu}_e$ energy in units of 15 MeV. The spectrum-dependent factor in the expression is $\approx 1.8$ for the simulation in Ref.\cite{Sagert:2008ka}. The signal rate has to be compared with the stochastic background rate $R_{0}= 2.20 \times 10^{3}$~bin$^{-1}$, with a rms fluctuation of 47 bin$^{-1}$. The solid (blue) line in fig.~\ref{IceCube} shows the event rate due to the ${\bar\nu}_e$ burst in IceCube for normal mass hierarchy [Eq.~(\ref{eq:normal})] in 2 ms bins. The dashed (green) line shows the expected background. In the narrow time window ($\sim 4$~ms) where the burst would show up, one would collect $\sim 2 \times 10^{4}$ events, as opposed to $\sim 8000$ without the QCD phase transition.

The vertical (red) band corresponds to the range of allowed signals for other possible oscillation scenarios. If we take the most pessimistic possibility that the oscillation effects are somehow arranged to produce the maximum possible signal outside the burst region and the minimum possible inside it, we get $\sim 10^{4}$ additional events in the burst bins over a steady background of $\sim 10^{4}$ SN events producing an excess with a significance of $\sim 100 \sigma$. The significance goes down at most quadratically with the SN distance, and even at $50$ kpc (the LMC) remains significant. The signature of the ${\bar\nu}_e$ burst, associated with the QCD phase transition in the SN core, would be spectacular in IceCube.

Since the temporal position of the $\bar{\nu}_e$ burst depends on the equation of state and the critical conditions for the phase transition, reconstructing the post-bounce time of the $\bar{\nu}_e$ burst is crucial. IceCube can reconstruct the bounce time to within $\pm 3.5$ ms at 95\% CL for a SN at 10 kpc~\cite{Pagliaroli:2009qy,Halzen:2009sm}, hence the post-bounce time of the QCD burst may be measured to within about $\pm 4$ ms, which is sufficiently accurate to discriminate between different model predictions (see Table I of Ref.~\cite{Sagert:2008ka}). 

\subsubsection*{Sub-Sampling at IceCube}

\begin{figure}[t]
\centering
\epsfig{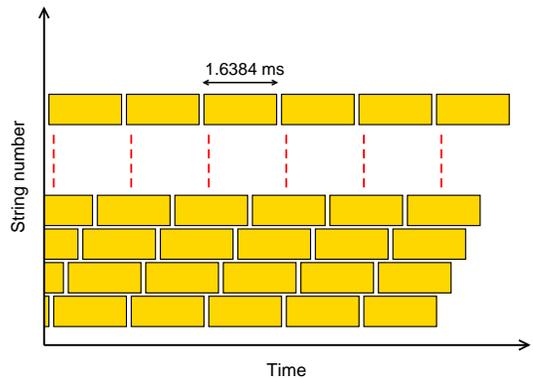}
\caption{Schematic representation of the relative time-offset 
between IceCube strings. 
\label{sampling}} 
\end{figure}

IceCube is usually quoted to have a time resolution of 1.6384 ms. This is indeed the time-window over which each optical module collects its data and transmits it to the data acquisition. The time synchronization of each module can however be controlled within $\lesssim 3$ ns relative to each other~\cite{privatecomm1}. Thus the intrinsic timing of IceCube is much better. The coarser time resolution arises only due to the 1.6384 ms binning. This can be in principle remedied, if one staggers the signal collection windows. In Fig.~\ref{sampling}, we schematically show the strategy. One could synchronize all modules on each string, but allow different strings to be offset in time by some fraction of the bin-size. For a sharp transient signal, different strings then attach a different time-stamp to the signal, and one would get a better time-resolution of the signal from a systematic comparison~\cite{privatecomm2}.

This procedure would not be very useful for weak signals where one loses significance rapidly by sub-dividing the detector. However, the QCD burst signal is expected to be quite large, and this strategy will allow a precise measurement of the rise-time of the signal.

\subsection{Detection at Super-Kamiokande}

\begin{figure}[t]
\centering
\epsfig{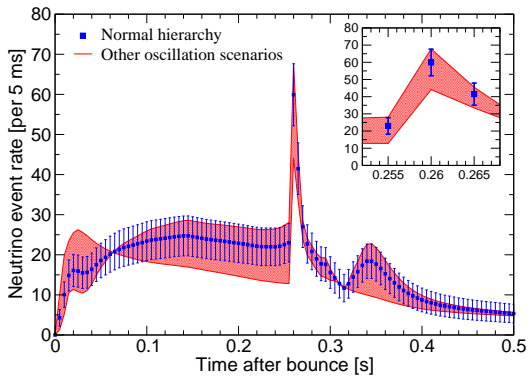}
\caption{Reconstructed neutrino event rates (note the subtle difference from Fig.~\ref{IceCube}) at Super-Kamiokande, for the neutrino emission shown in Fig.~\ref{simulation}. The inset shows the second burst blown up, in the same axis units.
\label{SK}} 
\end{figure}

In the following paragraphs we investigate the detection at Super-Kamiokande. The simplest procedure to perform is to similarly time-bin the events. We use bins of 5~ms, which is comparable to the duration of the second burst. Fig.~\ref{SK} shows the reconstructed neutrino event rates expected in Super-Kamiokande for the normal mass hierarchy in square points (blue) with statistical error bars. The range of allowed rates for other possible oscillation scenarios is shown by the shaded band (red). The QCD burst would produce $\sim 60$ excess events in the two relevant bins. The statistics are encouraging enough that the QCD-burst would be identifiable even at a SN distance of 20 kpc, which would contain $\sim 90 \%$ of the Galactic SNe~\cite{Mirizzi:2006xx}. However, the smaller the signal to noise ratio, the larger the risk that fluctuations in the accretion flow, which also produce variations in the electron flavor neutrino luminosities (see, e.g.,~\cite{Marek:2008qi}) can obscure a clear signal of the phase transition.

Super-Kamiokande's energy resolution can reveal a second distinguishing feature of the QCD phase transition burst -- the increase of the average neutrino energies. In particular, the largest energy increase is seen in the $\bar{\nu}_x$, which rapidly increases from 20 MeV to 35 MeV (see Fig.~\ref{simulation}). For the inverted mass hierarchy with large $\theta_{13}$, this energy increase would appear in the observed $\bar{\nu}_e$. Super-Kamiokande's energy resolution for positrons in this energy range is 10\% or better~\cite{Nakahata:1998pz}. The $\bar{\nu}_e$ energy can be determined from the positron energy, and hence the energy resolution of Super-Kamiokande permits the detection of this secondary effect due to the sudden compactification of the progenitor core during the quark-hadron phase transition. A third feature of the QCD phase transition is the transient increase in the $\bar{\nu}_e$ energies and luminosity, occurring $\sim 0.1$ s after the phase transition, as a result of accretion onto the remnant. This feature is most readily observable for the normal hierarchy.

\section{Discussion and Summary}

Core-collapse SNe can be used to probe the state of matter at high baryon densities and finite temperatures, beyond the reach of heavy-ion collision experiments at RHIC and the LHC, and complementary to those that will be studied with FAIR at the GSI Darmstadt. Using the MIT bag model, Ref.~\cite{Sagert:2008ka} investigated the quark-hadron phase transition during the early post-bounce phase of core-collapse SNe, demonstrating that it would cause a prominent burst of ${\bar\nu}_e$. We showed that this burst from the next Galactic core-collapse would be clearly detectable at IceCube and Super-Kamiokande, regardless of neutrino mixing scenarios. 

We have discussed three observational features of the QCD related $\bar{\nu}_e$ burst. The phase transition results in a clear and sharp rise in the signal rates at neutrino detectors, with Ice-Cube providing accurate measurements of the luminosity and timing of the burst. Super-Kamiokande also measures the energies of the neutrinos, revealing the increase in average neutrino energies predicted to occur during the burst. Finally, fluctuations in the accretion rate after the phase transition are also detectable. The accurately measured energetics and timing would reveal information about the critical conditions for phase transitions in the protoneutron star.

However, other transitions may occur in the extreme densities of the neutron star, {\it e.g.}~pion and kaon condensations \cite{transitions}. Indeed, any phase transition taking place in the early protoneutron star phase, through an extended mixed phase, may produce signals similar to those studied by Ref.~\cite{Sagert:2008ka} and this paper. Therefore, the detection of a burst of ${\bar\nu}_e$ during the early phase of a SN would provide a strong signature of a phase transition, and constrain the required conditions for its onset. The energetics and timing of the $\bar{\nu}_e$ burst would help observationally constrain the conditions required for the phase transition. In this regard we note that the physics of the phase transitions and whether a sharp ${\bar\nu}_e$ burst can arise therein remain to be studied systematically.

In addition to SN physics, the short rise time of the burst, combined with high statistics and energy measurement capabilities of SuperKamiokande, would allow the neutrino mass to be limited by the energy-dependent delays in arrival times. A massive neutrino with mass $m$ and energy $E$, traveling a distance $D$, will experience a delay relative to a massless neutrino, of
\begin{equation}
\Delta t \approx 0.515 \frac{ (m /{\rm eV})^2 (D /{\rm 10 kpc})}{(E /{\rm MeV})^2} \, {\rm s}.
\end{equation}
Considering the detector response and statistical errors, the mass limit from the QCD burst would be $\lesssim 1$~eV, assuming a rise time of 1 ms and that the rise includes 100 events~\cite{Beacom:1998ya}. This limit would be comparable with the most optimistic forecast obtained through a future measurement of SN neutrinos in SuperKamiokande~\cite{Pagliaroli:2010ik}. However, the detection of the first few events plays a key role in obtaining a strong limit, implying a large fluctuation on the mass bound. The mass limit achievable with the QCD burst would also be competitive with the less direct cosmological limits on neutrino masses~\cite{Hannestad:2010yi}.

This antineutrino burst signal also has implications for coincidence measurements with gravity wave detectors. The exceptional timing~\cite{Halzen:2009sm} provided by the narrow QCD burst will improve comparison with gravity wave detectors, for the procedure suggested by~\cite{Pagliaroli:2009qy,Leonor:2010yp}. A clear separation of the different bounce, accretion, and cooling phases of the explosion signal will allow for more sophisticated analysis of the signal.

In conclusion, our study confirms the crucial role of astrophysical multi-messengers played by neutrinos during a stellar collapse. In particular, it is fascinating to realize that weakly-interacting particles, like neutrinos, could become a powerful tool to probe the nature of strong interactions in extreme conditions.

\begin{acknowledgments}   

We thank John Beacom, Francis Halzen and Georg Raffelt for useful discussions. B.D.~acknowledges support from the Deutsche Forschungsgemeinschaft under grant TR-27, and the Cluster of Excellence: Origin and Structure of the Universe. T.F.~and M.L.~are supported by the Swiss National Science Foundation grant. no.~PP002-124879/1 and 200020-122287 and by CompStar, a research networking program of the European Science Foundation. S.H.~thanks the hospitality of the Werner-Heisenberg-Institut, where part of this work took place. I.~S.~is supported by the Helmholtz Research School on Quark Matter. J.~S.~is supported by the Heidelberg Graduate School of Fundamental Physics.

\end{acknowledgments}   



\begin{thebibliography}{00}   

\bibitem{Malek:2002ns}
  M.~Malek {\it et al.}  [Super-Kamiokande Collaboration],
  Phys.\ Rev.\ Lett.\  {\bf 90}, 061101 (2003)

\bibitem{Ando:2004hc}
  S.~Ando and K.~Sato,
  New J.\ Phys.\  {\bf 6}, 170 (2004)

\bibitem{Horiuchi:2008jz}
  S.~Horiuchi, J.~F.~Beacom and E.~Dwek,
  Phys.\ Rev.\  D {\bf 79}, 083013 (2009)

\bibitem{Woosley:2002zz}
  S.~E.~Woosley, A.~Heger and T.~A.~Weaver,
  Rev.\ Mod.\ Phys.\  {\bf 74}, 1015 (2002).

\bibitem{Totani:1997vj}
  T.~Totani, K.~Sato, H.~E.~Dalhed and J.~R.~Wilson,
  Astrophys.\ J.\  {\bf 496}, 216 (1998).

\bibitem{Dighe:2003be}
  A.~S.~Dighe, M.~T.~Keil and G.~G.~Raffelt,
  JCAP {\bf 0306}, 005 (2003).

\bibitem{Halzen:2009sm}
  F.~Halzen and G.~G.~Raffelt,
  Phys.\ Rev.\  D {\bf 80}, 087301 (2009)

\bibitem{Dighe:2008dq}
  A.~Dighe,
  J.\ Phys.\ Conf.\ Ser.\  {\bf 136}, 022041 (2008)
  
\bibitem{Takahara:1988yd}
  M.~Takahara and K.~Sato,
  Prog.\ Theor.\ Phys.\  {\bf 80}, 861 (1988).

\bibitem{Migdal:1974aa}
  A.~B.~Migdal,
  Phys.\ Lett.\ B {\bf 45}, 448 (1974).

\bibitem{Drago:2008tb}
  A.~Drago, G.~Pagliara, G.~Pagliaroli, F.~L.~Villante and F.~Vissani,
  AIP Conf.\ Proc.\  {\bf 1056}, 256 (2008)

\bibitem{Pons:2001}
  J.~A.~Pons, A.~W.~Steiner, M.~Prakash and J.~M.~Lattimer,
  Phys.\ Rev.\ Lett.\  {\bf 86}, 5223 (2001).

\bibitem{Hirata:1988}
  K.~S.~Hirata {\it et al.},
  Phys.\ Rev.\  D {\bf 38}, 448 (1988).

\bibitem{Gentile:1993ma}
  N.~A.~Gentile {\it et al.},
  Astrophys.\ J.\  {\bf 414}, 701 (1993).
  
\bibitem{Thorsson:1993bu}
  V.~Thorsson, M.~Prakash and J.~M.~Lattimer,
  Nucl.\ Phys.\  A {\bf 572}, 693 (1994)
  [Erratum-ibid.\  A {\bf 574}, 851 (1994)]

\bibitem{Nakazato:2008su}
  K.~Nakazato, K.~Sumiyoshi and S.~Yamada,
  Phys.\ Rev.\  D {\bf 77}, 103006 (2008).

\bibitem{Sagert:2008ka}
  I.~Sagert {\it et al.},
  Phys.\ Rev.\ Lett.\  {\bf 102}, 081101 (2009).

\bibitem{Kachelriess:2004ds}
  M.~Kachelriess {\it et al.},
  Phys.\ Rev.\  D {\bf 71}, 063003 (2005).

\bibitem{Duan:2007sh}
  H.~Duan, G.~M.~Fuller, J.~Carlson and Y.~Z.~Qian,
  Phys.\ Rev.\ Lett.\  {\bf 100}, 021101 (2008).

\bibitem{Sigl:2009cw}
  G.~Sigl {\it et al.},
  Nucl.\ Phys.\ Proc.\ Suppl.\  {\bf 188}, 101 (2009).
 
\bibitem{Hannestad:2006nj}
  S.~Hannestad, G.~G.~Raffelt, G.~Sigl and Y.~Y.~Y.~Wong,
  Phys.\ Rev.\  D {\bf 74}, 105010 (2006)

\bibitem{Dighe:1999bi}
  A.~S.~Dighe and A.~Y.~Smirnov,
  Phys.\ Rev.\  D {\bf 62}, 033007 (2000).

\bibitem{Fogli:2008ig}
  G.~L.~Fogli {\it et al.},
  Phys.\ Rev.\  D {\bf 78}, 033010 (2008).

\bibitem{Pagliaroli:2009qy}
  G.~Pagliaroli, F.~Vissani, E.~Coccia and W.~Fulgione,
  Phys.\ Rev.\ Lett.\  {\bf 103}, 031102 (2009)
  
\bibitem{privatecomm1}
  F.~Halzen, private communication.  
  
\bibitem{privatecomm2}
  B.D.~and S.H.~acknowledge extensive discussions with J.~F.~Beacom.
    
\bibitem{Mirizzi:2006xx}
  A.~Mirizzi, G.~G.~Raffelt and P.~D.~Serpico,
  JCAP {\bf 0605}, 012 (2006)

\bibitem{Marek:2008qi}
  A.~Marek, H.~T.~Janka and E.~Mueller,
  arXiv:0808.4136 [astro-ph].

\bibitem{Nakahata:1998pz}
  M.~Nakahata {\it et al.}  [Super-Kamiokande Collaboration],
  Nucl.\ Instrum.\ Meth.\  A {\bf 421}, 113 (1999)

\bibitem{transitions}
  A.~Ramos, J.~Schaffner-Bielich and J.~Wambach,
  Lect.\ Notes Phys.\  {\bf 578}, 175 (2001)
    J.~Schaffner-Bielich,
  Nucl.\ Phys.\  A {\bf 804}, 309 (2008)

\bibitem{Beacom:1998ya}
  J.~F.~Beacom and P.~Vogel,
  Phys.\ Rev.\  D {\bf 58}, 053010 (1998)
  
\bibitem{Pagliaroli:2010ik}
  G.~Pagliaroli, F.~Rossi-Torres and F.~Vissani,
  arXiv:1002.3349 [hep-ph].

\bibitem{Hannestad:2010yi}
  S.~Hannestad, A.~Mirizzi, G.~G.~Raffelt and Y.~Y.~Y.~Wong,
  arXiv:1004.0695 [astro-ph.CO].

\bibitem{Leonor:2010yp}
  I.~Leonor {\it et al.},
  arXiv:1002.1511 [astro-ph.IM].
  
\end{thebibliography}
\end{document}